\documentstyle[aps,twocolumn]{revtex}

\newcommand{\beq}{\begin{equation}}
\newcommand{\eeq}{\end{equation}}
\newcommand{\bea}{\begin{eqnarray}}
\newcommand{\eea}{\end{eqnarray}}
\newcommand{\ba}{\begin{array}}
\newcommand{\ea}{\end{array}}
\newcommand{\bc}{\begin{center}}
\newcommand{\ec}{\end{center}}

\newcommand{\ie}{{\it i.e.}}
\newcommand{\eg}{{\it e.g.}}
\newcommand{\etal}{{\it et al.}}
\newcommand{\bml}{\begin{mathletters}}
\newcommand{\eml}{\end{mathletters}}
\newcommand{\lanl}{LANL e-print }
\newcommand{\commentout}[1]{{}}
\newcommand{\half}{\hbox{$1\over2$}}

\newcommand{\HD}{{\cal H}}
\newcommand{\D}{{\cal D}}

\newcommand{\phidag}{\phi^\dagger}
\newcommand{\psidag}{\psi^\dagger}

\newcommand{\r}{{\bf r}}

\begin{document}
\draft
\wideabs{
\title{Coherent Control of Superfluidity in a Fermi Gas of Atoms}
%\title{Photoassociation-Induced Superfluidity in a Dilute Fermi Gas of Neutral Atoms}
\author{Matt Mackie$^*$, Jyrki Piilo$^*$, and Kalle-Antti Suominen$^{*\,\dagger}$}
\address{
  $^*$Helsinki Institute of Physics, PL 64, FIN-00014 Helsingin yliopisto, Finland\\
  $^\dagger$Department of Applied Physics, University of Turku,
    FIN-20014 Turun yliopisto, Finland}
\author{Juha Javanainen}
%\author{Robin C\^{o}t\'{e} and Juha Javanainen}
\address{Department of Physics, University of Connecticut, Storrs,
  Connecticut 06269-3046, USA}
\date{\today}
\maketitle

\begin{abstract}
We theoretically examine photoassociation of a two-component Fermi degenerate gas, focusing on
light-induced atom-atom interactions as a means to raise the
critical temperature of the BCS transition to a superfluid state. As it stands,
photoassociation-induced superfluidity is limited by spontaneous decay to experimentally
inconvenient light intensities {[}$\,$Mackie \etal, Opt. Express {\bf 8}, 118 (2000){]}.
We therefore propose to use coherent control in
photoassociation to a pair of molecular levels to cancel spontaneous emission, whereby the BCS
transition should occur within reach of current experimental techniques.
\end{abstract}
}

Studies of degeneracy in Fermi gases~\cite{FDGa,FDGa2} presently face
bottlenecks in reaching temperatures cold enough to form Cooper pairs. Below the Fermi
temperature, evaporative cooling of a dual Fermi gas begins to stall as
the near-unit-occupied low energy states exhibit Pauli blocking~\cite{PBLOCK}, while in a
mixture of Fermi and Bose gases the sympathetic cooling provided by ultracold bosons loses
efficiency when the bosonic heat capacity falls below its fermionic counterpart~\cite{FDGb}. The
lowest achieved temperatures are thus about a third of the Fermi temperature~\cite{PBLOCK,FDGb},
whereas Cooper pair formation requires a temperature of at least an order of
magnitude lower~\cite{LiBCSa,LiBCSb}. However, by adjusting atom-atom interactions, it could be
feasible to {\it raise} the BCS transition temperature to an experimentally accessible regime.
Competing means of adjustment include the Feshbach resonance~\cite{ASLFRa,ASLFRb,ASLFRc},
rf microwave fields~\cite{ASLRF}, dc electric fields~\cite{ASLDC}, and
photoassociation~\cite{ASLPAa,ASLPAb,OPUSMAG}.

Mathematically identical~\cite{FBCOMP},
photoassociation~\cite{DSPA} and the Feshbach resonance~\cite{DSFRa,DSFRb} have recently
been proposed for driving superfluidity in a Fermi degenerate gas of
atoms. In either case, the molecular state is, unfortunately, a
serious liability. For the Feshbach resonance, the bound state lies very close to the dissociation
threshold, and the subsequent  sensitivity, \eg, to collisions, will limit the lifetime of the Cooper pair--
an issue not yet fully addressed. On the other hand, photoassociation generally occurs to an excited
electronic state, and the superfluid lifetime is limited by spontaneous emission. As shown
previously~\cite{DSPA}, a ten second lifetime requires a far-far-off resonant photoassociation laser
(detuning $\sim 10^{14}\,$Hz) and, consequently, a critical temperature of a tenth the Fermi
temperature takes an enormous amount of light intensity ($\sim 10^8\,$W/cm$^2$). The purpose of
this Letter is therefore to develop a scheme for inducing the BCS transition to a superfluid state that is
both robust and user-friendly.

We model a binary mixture of fermionic atoms, denoted by the fields $\phi_{1,2}(\r)$,
photoassociating into two different bosonic molecules, denoted by $\psi_{1,2}(\r)$, a system that is
a neutral particle version of the boson-fermion model of high-temperature
superconductivity~\cite{HIGHTC}. Instead of opposite-spin electrons, the fermions herein would
typically be two states with different $z$ components of angular momentum in the same atom, which,
incidentally, bypasses the Pauli blocking of $s$-wave photoassociation. Now, a generic free-bound
transition {\it via} photon absorption leads to a molecule that is unstable against spontaneous
emission and, since there is no particular reason why this decay should deposit the ensuing population
back into the Fermi sea, such a molecule is considered lost forever. Consequently, assuming
photoassociation to vibrational levels in the same electronic manifold, we add a non-Hermitian term
proportional to the spontaneous decay rate of the excited electronic state $\gamma$, and incorporate
the possibility for quantum interference by considering a coherent superposition of molecular
amplitudes.

The Hamiltonian density for the atom-molecule system described above is
\bea
\hbar^{-1}\HD&=&\sum_l\left[\phidag_l\left(-{\hbar\nabla^2\over 2m}\right)\phi_l
  +\psidag_l\left(-{\hbar\nabla^2\over 4m} +\delta_l\right)\psi_l\right]
\nonumber\\
  &&-\sum_l \kappa_l(\r)\left(\psi_l^\dagger\phi_1\phi_2
    +\phi_2^\dagger\phi_1^\dagger\psi_l \right)
\nonumber \\
  &&-\half i\gamma\sum_{l,l'}\psidag_l\psi_{l'}
    +\lambda\phidag_2\phidag_1\phi_1\phi_2,
\label{HFULL}
\eea
The detuning of the photoassociating laser from the respective vibrational levels is
$\delta_l=\omega_\infty-\omega_L-\Delta_l$, where the binding energy of the $l$th molecular
state is $\hbar\Delta_l$, the energy of the photon is $\hbar\omega_L$, and
$\hbar\omega_\infty$ is the asymptotic energy difference between the two electronic manifolds
($l=1,2$).
A low-momentum approximation is implicit,
whereby relevant atom-atom collisions are described by a contact interaction of
strength $\lambda=4\pi\hbar a/m$, with $a$ being the $s$-wave scattering 
length. Similarly, correcting the
bosonic result~\cite{OPUSMAG} with a statistical factor of
$\sqrt{2}$, the (real) free-bound contact interaction strength $\kappa_l$ is given as
\beq
\kappa_l(\r)={\epsilon_R\lambdabar^{3/2}\over\sqrt{2}}\,
  \left[{I(\r)\over \left.I_0\right)_l}\right]^{1/2}.
\label{DL}
\eeq
Here $\epsilon_R=\hbar/2m\lambdabar^2$ is the usual photon recoil frequency, $2\pi\lambdabar$ is
the wavelength of the photoassociating light, and $I(\r)$ is the prevailing light intensity at the
position $\r$. Finally, if the
photoassociation rate coefficient $\alpha_l$ is known (in cm$^5$) at a temperature $T$ and
detuning $\delta$, the characteristic intensity $\left.I_0\right)_l$ is given (in
W/cm$^2$) as~\cite{OPUSMAG,DSPA}
\beq
\left.I_0\right)_l={\sqrt{\pi}\sqrt{\hbar\delta}c\hbar^4
  \over 2\alpha_lm^2 (k_BT)^{3/2}\lambdabar^2}\,e^{-\hbar\delta/k_BT}.
\eeq

\commentout{According to the Heisenberg equations of motion, the $l$th molecular field
evolves in time as
\beq
i\dot{\psi}_l=
  \left(-{\hbar\nabla^2\over 4m} +\delta_l\right) \psi_l 
    -\D_l\phi_1\phi_2-\half i\gamma\sum_{l'}\psi_{l'}.
\label{PSILDOT}
\eeq}

Assuming that $|\delta_l|$ is the largest frequency scale in the Heisenberg equations of motion, we
solve adiabatically for the molecular fields
$\psi_l$. Substituting the result into Eq.~(\ref{HFULL}) and keeping also the leading order of the
imaginary part in the energy, we obtain an effective Hamiltonian density involving only fermions
\bea
\hbar^{-1}\HD_{\text{eff}}&=&\sum_l\phidag_l\left(-{\hbar\nabla^2\over 2m}\right)\phi_l
  +\lambda_{\text{eff}}\,\phidag_2\phidag_1\phi_1\phi_2.
\label{HEFF}
\eea
The influence of photoassociating light on atom-atom interactions is now evident in the effective
collisional interaction strength
\beq
\lambda_{\text{eff}}=\lambda 
  -\left({\kappa_1^2\over\delta_1} +{\kappa_2^2\over\delta_2}\right)
    -\half i\gamma\left({\kappa_1\over\delta_1} +{\kappa_2\over\delta_2}\right)^2.
\label{LEFF}
\eeq

From the imaginary term in Eq.~(\ref{LEFF}), it is clear that spontaneous decay of (virtual) excited
molecules will limit the lifetime of the superfluid state through inelastic atom-atom scattering events.
Combining Eqs.~(\ref{DL}) and~(\ref{LEFF}) gives the radiative Cooper pair lifetime as
\beq
\epsilon_R\tau={2\over\rho\lambdabar^3}{\delta_1^2\over\gamma\epsilon_R}
    {1\over (1+R)^2},
\label{TAU}
\eeq
where the $R=(\delta_1/\delta_2)\sqrt{\left.I_0\right)_2/\left.I_0\right)_1}$ and $(\rho/2)^2$ was
used for the dual-atom density term
$\phidag_2\phidag_1\phi_1\phi_2$. We are evidently free to adjust the lifetime according to the
ratio $R$ and, in particular, $\tau=\infty$ is achieved by choosing $R=-1$. For
$\delta_2=\delta_1+\omega_{21}$, where $\hbar\omega_{21}=\hbar(\Delta_1-\Delta_2)>0$ is the
separation in energy between the molecular levels, and a characteristic intensity that scales with
binding energy as~\cite{OPUSMAG}
$\left.I_0\right)_2/\left.I_0\right)_1=\sqrt{\Delta_2/\Delta_1}$, we find
$\delta_1=-\omega_{21}/(1+\sqrt[4]{\Delta_1/\Delta_2})\approx
-\half\omega_{21}(1-\omega_{21}/\Delta_1)$, \ie, the photoassociating laser should be tuned
roughly halfway between the molecular levels. By exerting coherent control over photoassociation,
not only is spontaneous decay hereafter a non-issue, but the detuning and light intensity will assume
reasonable values.

Turning to the sought-after increase in the BCS transition temperature, we ignore
the native scattering length $a$ on the assumption that the associated collisional interaction alone
is too weak for experimental utility. The atom-atom interactions are now due solely to the
light-induced scattering length
\beq
{a_L\over\lambdabar}=-{1\over8\pi}\,{\epsilon_R\over\delta_1}\,{I\over \left.I_0\right)_1}\,
  \left(1+\sqrt[4]{\Delta_1\over\Delta_2}\,R\right).
\label{AL}
\eeq
Canceling spontaneous decay as above yields
$(1+\sqrt[4]{\Delta_1/\Delta_2}\,R)/\delta_1>0$, and therefore the attractive interaction
necessary for Cooper pairing. Having assumed the detuning is large enough to allow for adiabatic
elimination of the molecular field, the rigorous Fermi-Bose thermodynamics will reduce to the usual
BCS theory~\cite{DSFRb}; hence, the critical temperature for the superfluid transition is
approximately
$T_c=T_F\exp[-\half\pi/k_F|a_L|\,]$, where $T_F=\hbar^2k_F^2/2m k_B$ and
$k_F=(3\pi^2\rho)^{1/3}$ are the Fermi temperature and wave number, respectively.  Substituting
Eq.~(\ref{AL}) along with $R=-1$ gives
\beq
{T_c\over T_F}=\exp\left[-{18.1239\over(\rho\lambdabar^3)^{1/3}}\,
  {|\delta_1|\over\epsilon_R}\,{\left.I_0\right)_1\over I}\,
    \left|1-\sqrt[4]{\Delta_1\over\Delta_2}\right|^{-1}\right].
\eeq

Degeneracy has been observed for $^{40}$K~\cite{FDGa} and
$^6$Li~\cite{FDGb,FDGa2}, but the spectroscopic data are more plentiful for $^6$Li and we discuss
explicit numbers for this species only. In our example, we use the vibrational states $\nu=79$ and
$\nu=80$, with respective dissociation energies~\cite{NUMBERS} $\Delta_1=31.58$ and
$\Delta_2=22.61
\times 2\pi$ GHz, which leads to the detuning $|\delta_1|=4.3 \times 2\pi$ GHz. The fixed parameters
have the following values~\cite{OPUSMAG}: the wavelength is $2\pi\lambdabar=671$ nm; the
photon recoil frequency is $\epsilon_R=63.3\times 2\pi$ MHz; the density is assumed such that
$\rho\lambdabar^3=1$; and, finally, the characteristic intensity for the $\nu=79$ state is 9.8
mW/cm$^2$. Hence, the estimated intensity required to make $T_c/T_F=0.1$ is $I=30$
kW/cm$^2$, which is marginal at best.

To make further progress, we must appreciate the physics behind the canceling of spontaneous
emission. Let us regard the transitions from an atom pair to molecular levels 1 and 2 as charged
harmonic oscillators excited by light. The amplitudes of the oscillators may be adjusted by tuning the
laser frequency. When the light is tuned between the two transition frequencies, it also drives one of
the oscillators below resonance and the other one above resonance. The oscillations, and the radiation
from the two oscillators, therefore have opposite phases. By a suitable choice of the amplitudes (\ie,
light frequency) the spontaneous emission cancels. Analogous cancellations are of course well known
in quantum-optical few level atomic systems~\cite{FLCCa,FLCCb}. Although a molecular system
has many more degrees of freedom, the proliferation of rovibrational states-- and even
dissociation continua-- does not in principle invalidate our scheme. Despite a large number of
oscillators, there still exists a driving frequency that results in cancellation of the
radiation~\cite{NOTE}. As the oscillators closest to resonance have the largest amplitudes, such a
frequency should be close to the value deduced for the oscillators nearest to resonance.

Meanwhile, the change in the scattering
length reflects light shifts of the molecular levels. In the model with two molecular levels
of Eq.~(\ref{LEFF}), the two light shift terms proportional to $1/\delta_l$  work against each
other. At the point of vanishing spontaneous emission there is a nonzero change in the scattering
length, and of the right sign, only because the characteristic intensities (dipole matrix elements for
photoassociation) were taken to decrease (increase) going up the vibrational ladder, which is the
usual case in a molecule. Among the further off-resonance states, we expect the ones above resonance
dominate. The resulting added change in the scattering length then has the right sign for Cooper
pairing, and may be substantial since the light shift from a level decreases only inversely proportional
to the detuning.

With these considerations in mind, the multitude of states in a molecule actually helps our cause. For
instance, if we were to take into account in our argument four additional states below the state   
$\nu=79$ and four states above the state $\nu=80$ (see Ref.~\cite{NUMBERS} for the dissociation
energies), and assume the $\sqrt{\Delta}$ scaling of the characteristic intensities, we find that the
detuning needed to cancel spontaneous emission becomes $|\delta_1|=2.8 \times 2\pi$ GHz, and the
intensity to make $T_c/T_F=0.1$ is now $I =4$ kW/cm$^2$. This value is indeed an experimentally
feasible.

A truly quantitative prediction of
the intensity required for the BCS transition amounts to a calculation of an off-resonance light shift, a
notoriously difficult assignment. In this task experimental trial and error may be an easier way to
proceed. Should the dipole matrix elements turn out to be uncooperatively small, one may also
envisage a two-color scheme where the primarily photoassociated molecules are coupled with second
laser to another molecular state. Cancellation of spontaneous decay occurs (to lowest nontrivial order)
as in the one-color case, and the added shifts from off-resonant two-photon transitions will give a
handle to control atom loss and atom-atom interactions more independently than is possible with just
one laser.

All told, by allowing for the cancellation of spontaneous decay for reasonable values of detuning and
laser intensity, coherently controlled photoassociation should provide a means for creating a
superfluid state in a Fermi gas of atoms that is well within the reach of current experimental
techniques.

We thank John Calsamiglia and Chris Pethick for helpful discussions. This work
supported by the Academy of Finland (MM, JP, and KAS), as well as NSF and NASA (JJ). MM
thanks NORDITA Institute for Theoretical Physics for generous hospitality. JP acknowledges
support from the National Graduate School on Modern Optics and Photonics


\begin{references}
\bibitem{NOTE}
Normally, this kind of many-level cancellations of 
spontaneous-emission fail on angular momenta: Radiation from two 
oscillators that have different polarizations (say, one going in the 
x direction and the other in the y direction) cannot cancel. However, 
to zeroth order, vibrational level structure makes replicas of the 
same electronic transitions separated by vibrational level spacings, 
and the cancellation should indeed work.
\end{references}
\end{document}